\documentclass[12pt]{article}
\usepackage{epsf,cite}
\newcommand{\be}{\begin{equation}}
\newcommand{\ee}{\end{equation}}
\newcommand{\beqn}{\begin{eqnarray}}
\newcommand{\eeqn}{\end{eqnarray}}

\begin{document}

\title{\large{\bf A Note on Fermion and Gauge Couplings in Field Theory 
Models for Tachyon Condensation}}
\vskip 1.5cm
\author{ Dileep P. Jatkar\footnote{email: dileep@mri.ernet.in}, 
Subrata Sur\footnote{email: subrata@mri.ernet.in}\, and Radhika 
Vathsan\footnote{email: radhika@mri.ernet.in}\\
\\
{\it Harish-Chandra Research Institute,}\\ 
{\it Chhatnag Road, Jhusi, Allahabad 211 019, INDIA}\\
}
\maketitle

\vskip -7.8cm
{}~ \hfill\vbox{\hbox{hep-th/0107075}\hbox{MRI-P-010701} }%

\vskip 7.8cm

\begin{abstract}
{We study soliton solutions in supersymmetric scalar field theory with
a class of potentials. We study both bosonic and fermionic
zero-modes around the soliton solution. We study two possible couplings of
gauge fields to these models. While the Born-Infeld like coupling
has one normalizable mode (the zero mode), the other kind of coupling has 
no normalizable modes. We show that quantum mechanical problem
which determines the spectrum of fluctuation modes of the scalar, fermion 
and the gauge field is identical. We also show that only the lowest
lying mode, i.e., the zero mode, is normalizable and the rest of the
spectrum is continuous.}
\end{abstract}
\newpage

\section{Introduction}

The study of field theory models of tachyon dynamics has proved useful
in understanding the dynamics and stability of non-BPS D-branes, in
the light of the Sen conjectures\cite{sen1,sen2,sen3,sen4} on tachyon
condensation in string theory
\cite{bsft1,bsft2,bsft3,bsft4,bsft5,GS,KMM,GhS}. Zwiebach\cite{Zwie}
and then Minahan and Zwiebach\cite{MZ} have studied some solvable toy
models containing one scalar field representing the tachyon, with
Lagrangian densities containing at most two derivatives of the field.
Their models support lump solutions, whose fluctuation spectra
contain an unstable (tachyonic) mode. Minahan and Zwiebach\cite{MZ1}
have also studied a two-derivative model for tachyon dynamics in
superstring theory. They have shown that this model has a {\em stable}
kink solution. 

In a recent paper\cite{JV}, we generalized the model of Minahan
and Zwiebach, where we considered potentials of the form 
\be
V (\phi)  =  A q^n \phi^2(-\ln{\phi})^n, \quad n>1, \label{pot}
\ee
where $\phi(x)$ is the scalar field, $A$ and $q$ are arbitrary parameters and 
$n$ is a positive
integer. These potentials are not bounded below for odd values of $n$.
However, the corresponding field theory was found to possess a stable
kink solution. In this paper, we will first look at models with even $n$. 
They have potentials (\ref{pot}) which are bounded below and, not
surprisingly, have stable kink solutions. We will show that these
models can be supersymmetrized by introducing a fermionic field as
a supersymmetric partner of the original scalar field. The derivation
of bosonic fluctuation modes of the kink, which was studied in
\cite{JV} for $n$ odd, is equally valid for $n$ even. The functional 
form of the fermionic zero mode turns out to be similar. We then introduce 
a gauge field into the model.  This is equally applicable for both even 
and odd values of $n$. Minahan and Zwiebach have shown that there exists 
more than one way of coupling the gauge field. We will show that only 
the Born-Infeld like coupling has a normalizable zero mode. We also show 
that the ground state wavefunction of the underlying quantum mechanical 
problem is the only normalizable wavefunction. The rest of the spectrum is
continuous. Since the fluctuation spectrum of bosons as well as fermions is
governed by the same quantum mechanical problem, we find that
fluctuations of all these fields, for any $n\ge2$, contain only one
normalizable mode.

\section{The Supersymmetric Soliton}

The field theory model considered in \cite{JV} has the Lagrangian
density
\be
{\mathcal{L}} = - \frac{1}{2} \partial_{\mu}\phi\partial^{\mu}\phi 
              - V(\phi),
\ee
where $V(\phi)$ is given by Eq.~(\ref{pot}). For any positive integer $n$ 
this model has a soliton solution of the form
\renewcommand{\arraystretch}{1.5}
\beqn
\bar{\phi}(x) = \left\{ \begin{array}{ll}
              \exp{(e^{-\alpha_2 x})} 
             &      \mbox{ for $n = 2$}\\
             \exp{\left(- \alpha_n (x^2)^{-\frac{1}{n-2}}\right)}
             &      \mbox{  for $ n > 2$ } 
                       \end{array} \right. , \label{kink}
\eeqn
where
\be
 \quad \alpha_n = \left[\left(\frac{n}{2} - 1\right)^2 
              2 A q^n \right]^{-1/(n-2)}.
\ee
\begin{figure}[t]
\centerline{\epsfxsize 14.5cm
            \epsfbox{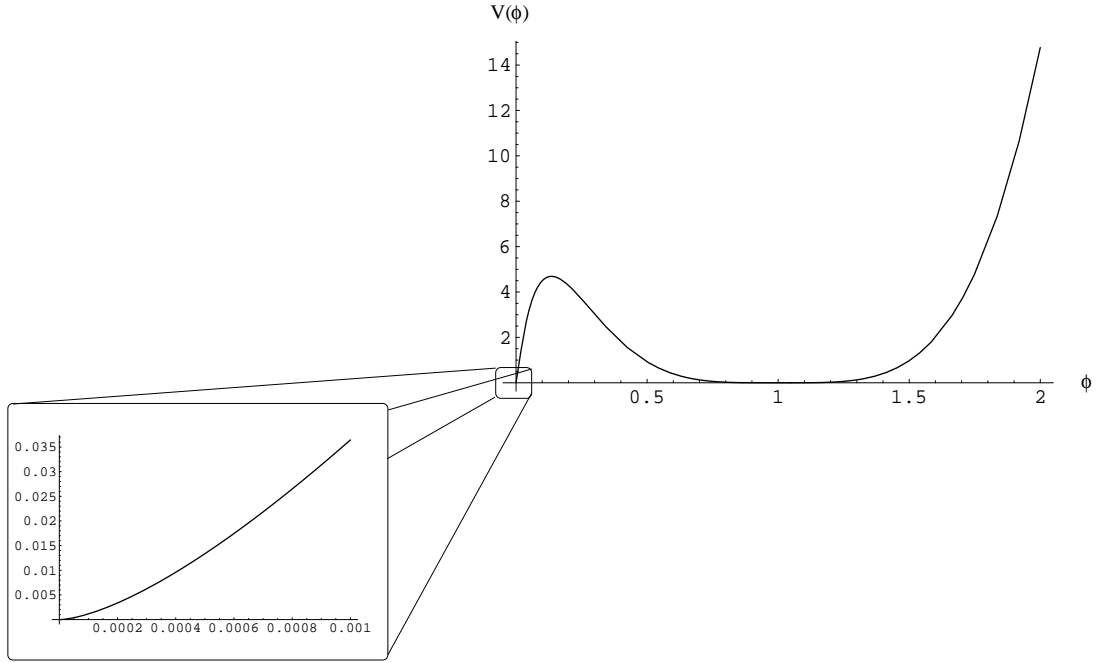}}
\caption{\small Generic form of $V(\phi)$ for even $n$. The enlarged
region shows the behaviour near the origin.}\label{fig-pot}
\end{figure}
While the potential (\ref{pot}) is not bounded below for odd $n$, 
for even $n$, it is. The generic form of the potential for even $n$ is
shown in Fig~\ref{fig-pot}.  In either case we have a local minimum at $\phi=0$ and a
maximum at $\phi = \exp{(-n/2)}$. The value of the potential at the maximum is
\be
V_{\rm max}= Aq^n {n^n\over 2^n}\; e^{-n}
\ee
At $\phi=1$, $n-1$ derivatives of the
potential vanish. For $n$ odd, this is a point of inflection and the potential 
becomes negative for $\phi >1$. On the other hand, for $n$ even, this is 
another minimum of the potential which is degenerate to the one at $\phi=0$.  
Due to the point of inflection at $\phi =1$ in the odd $n$ case, the kink 
solution turns out to be stable. The functional form of the kink
solution as well as  the quantum mechanical problem for the fluctuation 
spectrum is the same for the even $n$ case. As mentioned earlier, the
potential is bounded for even $n$ and it is possible to write down a
supersymmetric extension to the corresponding bosonic field theory
model.

\subsection{Mass of the Soliton} 

The mass of the soliton can be calculated from the following expression
for the total energy of the field configuration $\bar \phi(x)$:
\be
 E_{\rm sol} = \int_{0}^{\infty}dx \left[ \frac{1}{2}(\partial_{x}
 \bar{\phi})^2 + V(\bar{\phi}) \right]
\ee
Using the equation of motion $\partial_{x}\bar{\phi} =
\sqrt{2V(\bar{\phi})}$ and substituting from Eq,~(\ref{kink}) 
for $\bar{\phi}(x)$,
\beqn
E_{\rm sol} = \sqrt{A q^n 2^{3-n}} \,\, \Gamma(n/2 + 1). 
\eeqn
The ratio of the soliton mass to the tension of the original brane,
which is same as the value of the potential at the maximum, is
\be
{E_{\rm sol}\over V_{\rm max}} = {e^{n}\Gamma({n\over 2}+1)\over 
2 \sqrt{Aq^n2^n}n^n}
\ee 
It is straightforward to see that this result agrees with that obtained
by Minahan and Zwiebach for $n=1$, {\em viz.} $e/\sqrt{2\pi}$ 
(with $A=1/4$ and $q=2$). For other $n$, the ratio is considerably 
smaller than $1$. 

\subsection{Coupling to Fermions}

In this section, we will study a supersymmetric model with the bosonic
potential given in Eq.~({\ref{pot}) for $n$ even. Consider a
Lagrangian density with a fermion field $\Psi(x)$ coupled to the scalar
field $\phi$, 
\beqn
 {\mathcal{L}} = -\frac{1}{2} \partial_{\mu}\phi\partial^{\mu}\phi
 - {1\over 2} {\cal V}^2(\phi)  +\bar{\Psi} i \partial\!\!\!/\Psi + \bar{\Psi}
\Psi \frac{d {\cal V}(\phi)}{d\phi},
\eeqn
where
\be
{\cal V}(\phi) = \sqrt{2 V(\phi)} = \sqrt{2A} q^{n/2} \phi (-\ln{\phi})^{n/2}.
\ee
where $n$ is now a positive even integer. $\Psi(t,x)$ is a two component 
Majorana spinor. The equations of motion are
\beqn
\partial_{\mu}\phi\partial^{\mu}\phi - {\cal V}(\phi){d{\cal V}\over d\phi} + 
\bar\Psi\Psi{d^2{\cal V}\over d\phi^2} &=& 0, \\
i\partial\!\!\!/\Psi + \Psi \frac{d {\cal V}}{d\phi} &=& 0.
\eeqn
A classical solution to these equations of motion is obtained by setting the
fermion field $\Psi$ to zero and then solving the pure bosonic problem. The
equation of motion thus obtained was solved in \cite{JV} to get the  soliton
solution $\bar{\phi}(x)$ of Eq.~(\ref{kink}). Within the bosonic sector, the 
small fluctuation spectrum around the soliton solution was studied in \cite{JV}.
In particular, the bosonic zero mode was calculated exactly: 
\be
\psi_0(x) = x^{-n/(n-2)} \, \exp{ \left( -\alpha_n x^{-2/(n-2)}
\right)}.\label{gs}
\ee
Here we will study the fermionic zero mode in the background of the
soliton (\ref{kink}). This is obtained by solving for the zero mode 
$\Psi^{(0)}$ of the spinor equation with a potential $ U^{\prime}(\bar{\phi})$.
Since the equation is linear in $\Psi$, the zero mode will essentially satisfy
the same equation of motion. The two components $\Psi_{1,2}^{(0)}(x)$
of the zero mode satisfy the equations
\beqn
 \mp\partial_x\Psi_{1,2}^{(0)} +{\cal V}^{\prime}(\bar{\phi})\Psi_{1,2}^{(0)}
= 0,
\eeqn
the general solutions to which are 
\be
 \Psi_{1,2}^{(0)}(x) = {x}^{\mp\frac{n}{(n-2)}}\exp{\left(\mp
{\alpha_n} x^{-\frac{2}{n-2}}\right)}.
\ee
Of these two solution, only $\Psi_{1}^{(0)}$, corresponding to the
`$-$' sign, is well behaved when $x\to0$ and $x\to\infty$ as can be 
checked easily.  So the normalizable fermionic zero mode 
is given by 
\renewcommand{\arraystretch}{1.6}
\beqn
\Psi^{(0)}_1(x)& =& \left\{  \begin{array}{ll}
  \exp\left[  e^ {-\alpha_2 x} - \alpha_2 x \right]
  & n=2 \\
 {x}^{-\frac{n}{(n-2)}}\exp{(-\alpha_n 
           x^{\frac{-2}{n-2}})}          
  & n > 2                       
 \end{array}\right., \nonumber\\
\Psi^{(0)}_2(x) &=& 0.
\eeqn
Thus wecan see that the soliton breaks half the supersymmetry of the original
model. Notice, the wavefunction $\Psi^{(0)}_1(x)$ is identical to the
bosonic zero mode (\ref{gs}). This is expected since $\Psi$ is a
superpartner of $\phi$. Hence, the quantum mechanical
problem governing the two is identical. We will encounter this quantum
mechanical problem again when we discuss gauge field couplings.

\section{Coupling Gauge Fields}

Minahan and Zwiebach \cite{MZ2} have considered two possible ways of
introducing gauge fields in these kinds of models. We consider both
of them below and study fluctuations of the gauge field about the brane 
(soliton). 

We first consider the Born-Infeld action for a D-p-brane coupled to a 
gauge field $A_{\mu}$:
\beqn
S_{BI} =  - {\mathcal T} \int dt d^{p+1}x \, V(T)
\sqrt{-\det(\eta_{\mu\nu} + F_{\mu\nu})}, \label{BI}
\eeqn
where $V(T)$ is the potential for the tachyon field $T(x)$.  We are
studying tachyon condensation only along one direction, say $x$ (all
other $p$ directions will be referred to as $y$), and its potential is
derived from our field theory potential of Eq.~(\ref{pot}). Consider a
field redefinition (as in \cite{JV} following \cite{GS}) 
\be
\phi = e^{-T} \; \Rightarrow \; V(T) = Aq^nT^n\,  e^{-2T}. \label{twentyone}
\ee
Since we wish to study  small fluctuations of the gauge field around
the soliton solution, we can make the approximation
\be
\sqrt{-\det(\eta_{\mu\nu} + F_{\mu\nu})} \approx 1 + {1\over 4}
F_{\mu\nu}F^{\mu\nu} + \cdots,
\ee
and retain terms upto quadratic order. The action involving the gauge 
and tachyon fields is then given by
\be
S = - {1\over 2}\int dt d^{p+1}x \,\, 
    e^{-2T} \left[ \partial_{\mu}T\partial^{\mu}T +
Aq^nT^n\left( 1 + {1\over 4}F_{\mu\nu}F^{\mu\nu}\right) \right]. 
\ee
We choose the axial gauge condition, {\em i.e.} $ A_x = 0$. This
is equivalent to setting $ F_{x\bar\mu} = \partial_x A_{\bar\mu}$,
where $\bar\mu$ represents the coordinates other than $x$.
To calculate the spectrum of gauge fluctuations about the kink, we
look at the gauge part of the action with $T(x) = \bar{T}(x) =
-\ln\bar\phi$.  At this point, it is useful to make a field redefinition
for the gauge fields:
\beqn
B_{\bar\mu} &\equiv& \sqrt{Aq^n/2} e^{-\bar{T}} \bar{T}^{n/2}
A_{\bar\mu},\\
F_{\bar\mu\bar\nu}F^{\bar\mu\bar\nu}& =& \partial_{\bar\mu} B_{\bar\nu}
                      -\partial_{\bar\nu} B_{\bar\mu}.
\eeqn
With this field redefinition and substitution of $T(x) = \bar{T}(x)$ in the 
action for the gauge field gives
\be
S(\bar T, B(A)) = - \int dt d^py dx \left[ {1\over4}
F_{\bar\mu\bar\nu}F^{\bar\mu\bar\nu} + B_{\bar\mu}{\mathcal
O}(x)B^{\bar\mu} \right]
\ee
where ${\mathcal O}(x)$ is the Schr\"{o}dinger operator governing 
gauge field fluctuations in the soliton background. It can be written
in terms of $\bar T$ as
\beqn
{\mathcal O}(x) &=& -{\partial^2 \over \partial x^2} + U(x)\nonumber
\\ 
&=& -{\partial^2 \over \partial x^2} 
+ \bar{T}^{\prime\prime}\left({n\over 2} \bar T^{-1} - 1\right)
+ (\bar{T}^{\prime})^2 \left({n\over 2} \bar T^{-1} - 1\right)^2
- {n\over 2}{(\bar{T}^{\prime})^2\over \bar{T}^2}. \label{schro}
\eeqn
It is clear from the field redefinition (\ref{twentyone}) that 
\be
\bar{T}(x) 
= \alpha_n x^{-{2 \over{n-2}}}.\label{tach} 
\ee
Substituting this in (\ref{schro}), we find that the potential $U(x)$ in 
the Schr\"{o}dinger operator is the same as that
for the scalar fluctuations about the kink solution obtained in
\cite{JV}:
\be
U(x) = \frac{2}{(n-2)^2} \left[
                2 \alpha_n^2 \left(\frac{1}{x^2}\right)^{n/(n-2)}
       \!\!\!\! -3n \alpha_n \left(\frac{1}{x^2}\right)^{(n-1)/(n-2)}
       \!\!\!\!   +n(n-1) \frac{1}{x^2} \right].\label{qmpot}
\ee
\begin{figure}[t]
\centerline{\epsfxsize 10cm
            \epsfbox{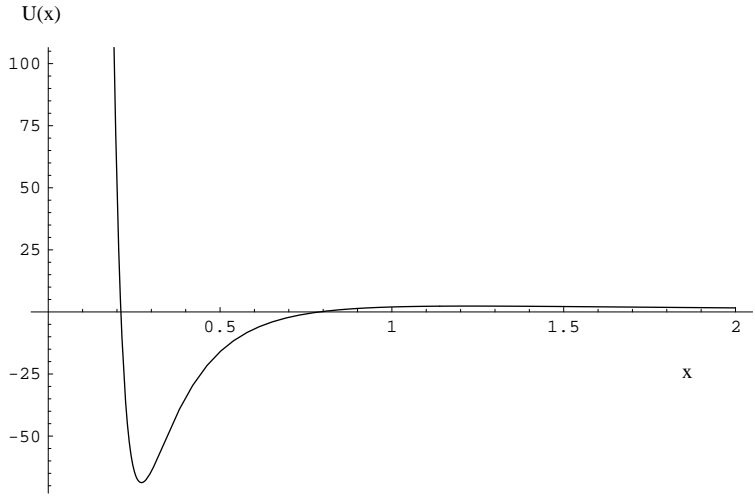}}
\caption{\small The potential for the Schr\"{o}dinger equation
governing the fluctuations}\label{fig-qmpot}
\end{figure}
We saw in \cite{JV} that this defines  a supersymmetric (SUSY) quantum
mechanics problem\cite{Khare} with superpotential given by
\beqn
W(x)=\frac{1}{n-2}\left[ \frac{n}{x} -
\frac{ 2 \alpha_n}{x^{n/(n-2)}}\right],
\eeqn
whose (massless) ground state can be calculated exactly.
The higher excited states can be calculated algebraically using SUSY
techniques if one can solve the Ricatti equation
\beqn
W^2 + W^{\prime} = \tilde{W}^2 - \tilde{W}^{\prime} + R 
\eeqn
where $\tilde{W}$ is another  superpotential and $R$ is some
constant\cite{Khare}.
In our case, we find $\tilde{W} = - W $ and $R=0$, which implies that
the ground state  of  the problem with potential $\tilde U = \tilde{W}^2 -
\tilde{W}^{\prime}$ is not normalizable. Thus it supports no bound state, and 
the entire spectrum is continuous. Using methods of SUSY quantum
mechanics we can relate the spectrum of $U$ with that of $\tilde U$. 
The continuous spectrum of $\tilde U$ implies that for the potential $U$ of
Eq.~(\ref{qmpot}), there exists only one bound state and that is the
zero-energy ground state calculated in \cite{JV}. This fact is also
clear from the shape of the potential (see Fig.(\ref{fig-qmpot})),
which approaches zero for large $x$. Using SUSY, we also
conclude that there is only one normalizable fermionic mode. 

A simpler gauge coupling than the Born-Infeld kind has also been considered
in \cite{MZ2}. The action looks like
\be
S^{\prime} = - {1\over 4}\int  dt d^{p+1}x \,\, e^{-2T} F_{\mu\nu}F^{\mu\nu}.
\ee
Once again choosing the axial gauge and using the following field redefinition
\be
B_{\bar\mu} = e^{-\bar T} A_{\bar\mu},
\ee
we find that the  gauge fluctuations are governed by a  
Schr\"{o}dinger operator
\be
{\mathcal O}(x) = - {{\partial^2} \over {\partial x^2}} + 
              (\bar T^{\prime})^2 - 2 \bar T^{\prime\prime}
\ee
The zero energy ground state eigenfunction of this operator can be 
expressed in terms of $\bar T$. Substituting the explicit form of $\bar T$ 
from Eq.~(\ref{tach}), it is evident that this eigenfunction is not
normalizable. Thus we find that gauge coupling of this type does not
lead to normalizable gauge fluctuations about the soliton. Apart from
the numerical factors the profile of the gauge zero mode is identical
to that of the kink, which means the zero mode fluctuations grow as
we go away from the core of the soliton. This feature is contrary to
that of the D-branes where one expects the gauge zero mode to be
localized near the core of the soliton. Thus this coupling is less
suitable compared to the Born-Infeld coupling. 

\section{Conclusions}

This paper is a continuation along the lines of investigation started
in \cite{JV} where we studied soliton solutions for a class of scalar field 
theory models with potentials that are not bounded below. They
correspond to odd values of a parameter $n$. In this
paper we study the same for potentials that are bounded below,
corresponding to even values of $n$.

We calculate the mass of the soliton and compare it to the tension of the
original D-brane. This calculation is valid for both odd and even
values of $n$. Our formula reproduces the result of Minahan and
Zwiebach \cite{MZ} for $n=1$. As $n$ increases, the mass decreases and
the ratio of mass to D-brane tension becomes considerably lesser than 1. 

The models with  $n$ even can be supersymmetrized, and accordingly, we add a
fermionic field to the model and study its fluctuations about the
soliton. We find one normalizable zero mode and show that it breaks half 
the supersymmetry. The wavefunction for this mode is identical to that of 
the scalar zero mode.

We then study two types of gauge couplings following the proposals of
Minahan and Zwiebach \cite{MZ2}. This analysis is also valid for any value
of $n$, even or odd. We find that for a Born-Infeld type of coupling,
the gauge fluctuations have only one normalizable mode, {\em viz.} the
zero mode. In fact, the quantum mechanical problem governing the gauge
fluctuations in this case turns out to be the same as that for the
scalar as well as the fermionic fluctuations about the soliton. Thus, 
the spectrum of fluctuations in all these fields have only one normalizable 
mode, which has zero energy. The rest of the fluctuation modes have
continuous spectrum.

We show that the other kind of gauge coupling has no normalizable
fluctuations about the soliton. The zero mode, which is also not 
normalizable, grows away from the core of the soliton. This is
contrary to what we know about the D-branes and tachyon condensation.
Even within field theory, growth of gauge fluctuations away from
the core of the soliton is counterintuitive.

Comparing contrasting features of these two possible gauge couplings, 
especially the behaviour of zero modes, we conclude that the
Born-Infeld type gauge coupling is more appropriate.

With the Born-Infeld type coupling of the gauge field to the scalar 
field and consequently to the soliton, it is then straightforward to 
obtain lower dimensional branes as well as intersecting brane 
configurations, using techniques studied in \cite{HH}, i.e., by 
turning on an electromagnetic field in different directions.

\vspace*{3mm}

\noindent{\bf Acknowledgements}:
We would like to thank Ashoke Sen for useful discussion.
R.V. acknowledges support by CSIR, India, under grant no. 
9/679(7)/2000-EMR-I.

\end{document}